\documentclass[aps,prfluids,preprint]{revtex4-1}

 \usepackage{epsfig}
 \usepackage{color}
 \usepackage{amsmath}
 \usepackage{amsthm}
 \usepackage{amsfonts}
 \usepackage{amssymb}
\usepackage{graphicx}
\usepackage{epstopdf}

\usepackage{ulem}
\usepackage[dvipsnames]{xcolor}

\begin{document}

\title{Emergence of skewed non-Gaussian distributions of velocity increments in isotropic turbulence}

\author{W. Sosa-Correa}
\affiliation{Laborat\'orio de F\'{\i}sica Te\'orica e Computacional, Departamento de F\'{\i}sica,
Universidade Federal de Pernambuco, 50670-901 Recife, Pernambuco, Brazil}

\author{R.~M. Pereira}
\affiliation{Laborat\'orio de F\'{\i}sica Te\'orica e Computacional, Departamento de F\'{\i}sica,
Universidade Federal de Pernambuco, 50670-901 Recife, Pernambuco, Brazil}

\author{A.~M.~S. Mac\^edo}
\affiliation{Laborat\'orio de F\'{\i}sica Te\'orica e Computacional, Departamento de F\'{\i}sica,
Universidade Federal de Pernambuco, 50670-901 Recife, Pernambuco, Brazil}

\author{E.~P. Raposo}
\affiliation{Laborat\'orio de F\'{\i}sica Te\'orica e Computacional, Departamento de F\'{\i}sica,
Universidade Federal de Pernambuco, 50670-901 Recife, Pernambuco, Brazil}

\author{D.~S.~P. Salazar}
\affiliation{Unidade de Educa\c{c}\~ao a Dist\^ancia e Tecnologia, Universidade Federal Rural de Pernambuco,
52171-900 Recife, Pernambuco, Brazil}

\author{G.~L. Vasconcelos}
\affiliation{Departamento de F\'{\i}sica,
Universidade Federal do Paran\'a, 81531-990 Curitiba, Paran\'a, Brazil}




\begin{abstract}
Skewness and non-Gaussian behavior are essential features of the distribution of short-scale velocity increments in isotropic turbulent flows. Yet, although the skewness  has been generally linked to time-reversal symmetry breaking and vortex stretching,  the form of the  asymmetric heavy tails remain elusive. Here we describe the emergence of both properties through an exactly solvable stochastic model with a scale hierarchy of energy transfer rates. From a statistical superposition of a local equilibrium  distribution weighted by a background density, the increments distribution is given by a novel class of skewed heavy-tailed distributions, written as a generalization of the Meijer $G$-functions. Excellent agreement in the multiscale scenario is found with numerical data of systems with different sizes and Reynolds numbers. Remarkably, the single scale limit provides poor fits to the background density, highlighting the central role of the multiscale mechanism. Our framework can be also applied to describe the challenging emergence of skewed distributions in complex systems.
\end{abstract}

\maketitle

\section{Introduction}

The phenomenon of turbulence is plentiful of challenging features that still remain elusive after decades of efforts~\cite{frisch,review}.
In particular, the negative skewness and non-Gaussian behavior of the distribution of velocity increments between close points
in a homogeneous and isotropic turbulent flow have long figured among the most intriguing ones.
Though the negative asymmetry can be derived from the Navier-Stokes equations and has been connected
to the time-reversal symmetry breaking~\cite{xuetal14},
elucidating its physical origins and determining the form of the heavy tails persist as long-standing open questions.

Indeed, understanding the statistical properties of velocity fluctuations has always been, and remains, an essential issue in turbulence.
A significant step in this direction was Kolmogorov's theory of turbulence~\cite{frisch}.  
One of its few exact results is the so-called 4/5-law:
$\langle (\delta v_r)^3\rangle=-\frac{4}{5} \langle \varepsilon\rangle r$,
where $\delta v_r=v(x+r)-v(x)$ represents the longitudinal velocity increment and $\langle \varepsilon\rangle$ is the mean energy dissipation rate.
For homogeneous and isotropic turbulent flows, in which $\langle \delta v_r\rangle=0$,
Kolmogorov's 4/5-law implies negative skewness and non-Gaussian statistics of velocity increments.
Considerable effort has been also devoted to investigate the scaling properties of higher-order structure functions,
$\langle (\delta v_r)^n\rangle\sim r^{\zeta_n}$, $n>3$, for which  no exact results are known~\cite{frisch,PRF2019}.
%
%
Moreover, a renewed interest has arisen as well in the study of the increments distribution itself,
rather than its set of moments~\cite{andrews_1989,Kailetal92,castaing_PhysD90,naert-1998,chevetal03,chevetal05, chevetal06}.
In particular, it has long been known that velocity increments for large separations  tend to be Gaussian distributed,
whereas non-Gaussian behavior is observed at short scales~\cite{frisch}.
In this context, a more recent work~\cite{pnas2014} found that short-scale non-Gaussian effects appear at Reynolds numbers
much smaller than initially thought.

Here we report on a statistical approach to the distribution of short-scale velocity increments
in isotropic turbulent flows that describes 
the emergence of both the negatively skewed asymmetry and non-Gaussian heavy tails, with very nice agreement with numerical turbulence data of systems featuring distinct sizes and Reynolds numbers.
Our work is based on two central tenets of turbulence theory~\cite{frisch,review}, namely the intermittency phenomenon and the concept of energy cascade,
whereby energy is transferred from large to small eddies until dissipation by viscous forces at the shortest (Kolmogorov) scale.

Our intermittency model is built upon a hierarchy of multiple coupled scales of {\it energy transfer rates}~\cite{pre2010,pre2017,pre2018}. 
The marginal distribution of short-scale velocity increments~$P(\delta v_r)$ 
is related to the energy transfer rate~$\varepsilon_\ell$ at a larger scale~$\ell$ [see Eq.~(\ref{eq:P1e}) for a formal definition of $\varepsilon_\ell$] 
through a statistical superposition of the conditional distribution~$P(\delta v_r|\varepsilon_\ell)$, weighted by a background distribution~$f(\varepsilon_\ell)$ obtained in exact closed form from our model.
By considering $P(\delta v_r|\varepsilon_\ell)$ as a Gaussian with nonzero mean  characterized by an asymmetry parameter $\mu$, we obtain an exact $P(\delta v_r)$ in the form
of a novel class of skewed functions with stretched exponential heavy tails.
These newly defined functions constitute a generalization of the Meijer-$G$ functions
and, to our knowledge, have never been 
considered in the literature. 

The theoretical predictions emerging from this multiscale scenario are found to be in excellent agreement
with turbulence data from two extensive and independent numerical simulations of the Navier-Stokes equations.
%
Remarkably, a  poor agreement is found if only a single scale is considered. 
%
Also, the origin of the stretched exponential heavy tails is  shown to be directly related to the multiscale behavior, since a simple exponential decay would result if only a single scale were present. Therefore, our results highlight the crucial role of the interplay of multiple coupled scales of energy transfer rates,  advancing on the multiscale modeling of turbulent systems in an alternative way to other approaches, such as multiplicative cascades \cite{frisch}, shell \cite{bif03} and Lagrangian \cite{bifetal07,JohMen17} models. Moreover, our framework can be also applied to investigate the  emergence of skewed distributions in other complex systems, such as financial markets~\cite{complex1} and biological systems~\cite{complex2}.


\section{Theoretical Background}\label{sec:theo}
\label{sec:theory}

We work under the formalism of a unified hierarchical approach to describe the statistics of fluctuations
in multiscale complex systems~\cite{pre2010,pre2017,pre2018}. 
This framework, called H-theory, is an extension to multiscale systems of the compounding~\cite{andrews_1989,castaing_PhysD90} or superstatistics~\cite{beck} approaches to describe complex fluctuating phenomena. In this formalism, 
the probability distribution of the relevant signal---say, the velocity increments---at short scales
is given by a statistical superposition of a large-scale conditional distribution weighted
by the distribution of certain internal degrees of freedom related to the slowly fluctuating environment,
\begin{align}
P(\delta v_r)=\int_0^\infty P(\delta v_r|\varepsilon_\ell)f(\varepsilon_\ell) d\varepsilon_\ell,
\label{eq:Px}
\end{align}
where the variable $\varepsilon_\ell$ characterizes the local equilibrium at scale $r$. The large-scale conditional distribution is assumed to be known, so that the complex statistical properties of the turbulent state are entirely captured by the weighting density~$f(\varepsilon_\ell)$,
which incorporates the effect of the fluctuating energy flux (intermittency).
In turbulence modelling, the conditional distribution $P(\delta v_r|\varepsilon_\ell)$ in Eq.~(\ref{eq:Px}) is often chosen to be a Gaussian with zero mean, while several different weighting distributions have been used, such as the gamma \cite{andrews_1989}, lognormal \cite{castaing_PhysD90,gagneetal94,chabaud_PRL94,naert-1998,yakhot2006} and inverse-gamma \cite{beck} distributions.
A distinctive feature of our formalism, however, is that the distribution $f(\varepsilon_\ell)$ in (\ref{eq:Px}) is not prescribed {\it a priori}---as in these previous works---, 
but rather is calculated from a hierarchical intermittency model; see below.

One important physical assumption built into Eq.~(\ref{eq:Px}) is the separation of time and length scales: the background variable  $\varepsilon_\ell$ is supposed to vary  more slowly (in time and space) than the signal $\delta v_r$~\cite{beck}, thus allowing it to reach a quasi-equilibrium distribution $P(\delta v_r|\varepsilon_\ell)$.
In the statistical mechanics language, structures of size $\ell$ act as a `heat bath' for the fast fluctuating quantity $\delta v_r$~\cite{pre2012}.
In the turbulence context, $\varepsilon_\ell$ can be  associated with  the energy transfer rate from scale~$\ell$ towards smaller scales, where $\ell\gg r$ in view of the assumed scale separation. 

Following \cite{gagneetal94}, we consider the energy transfer rate $\varepsilon_\ell$ at scale $\ell$ as defined by 
\begin{align}
\varepsilon_\ell(x)=15\nu\left[\frac{1}{\ell}\int_x^{x+\ell} \left(\frac{\partial v}{\partial x'}\right)^2dx'-\left(\frac{\delta v_\ell}{\ell}\right)^2\right],
\label{eq:vareps}
\end{align}
where $\nu$ is the viscosity. The first term in the right-hand side of (\ref{eq:vareps}) is the space average of the  dissipation rate over a volume of size $\ell$, which is Obukov's proposal for estimating the rate of energy transfer~\cite{frisch}, whereas the term $15\nu \left(\frac{\delta v_\ell}{\ell}\right)^2$ takes into account the energy dissipation  at the scale $\ell$ itself \cite{gagneetal94}. For large $\ell$ (say, in the inertial range), the second  term is negligible and so $\varepsilon_\ell$ agrees with  Obukov's prescription for the energy transfer rate.
In Ref.~\cite{gagneetal94} it is  argued that the energy transfer rate $\varepsilon_\ell$ defined in (\ref{eq:vareps}) can be approximated  by $\varepsilon_\ell \approx 15 \nu \epsilon_\ell /r^2$, where $\epsilon_\ell =\langle (\delta v_r)^2 \rangle  - \langle \delta v_r \rangle^2$ is the variance of $\delta v_r$ at the scale $\ell$, meaning that the averages $\langle (\cdots)\rangle$ are performed over  windows of size $\ell$.
Here we shall  make  a similar assumption and take the variance, $\epsilon_\ell$, of $\delta v_r$ over a region of size~$\ell$  as a proxy measure for the energy transfer rate~$\varepsilon_\ell$. 
We note, however, that in our approach the scale~$\ell$ is not initially known and must be determined from the velocity data, as explained in Section~\ref{sec:data}.

Experimental and theoretical studies on homogeneous and isotropic turbulent flows indicate~\cite{andrews_1989,castaing_PhysD90,naert-1998,chevetal05,gagneetal94,chabaud_PRL94,SKS}  
that the conditional distribution 
$P(\delta v_r|\epsilon_\ell)$ 
is given by a Gaussian with variance  $\epsilon_\ell$.
For the sake of simplicity, a Gaussian  with {\it zero} mean is often considered in theoretical turbulence models~\cite{andrews_1989,pre2010,pre2017,beck,pre2012}, leading to {\it symmetric}  (i.e.,~non-skewed) distributions $P(\delta v_r)$. 

Here we introduce a model for $P(\delta v_r|\epsilon_\ell)$ that yields an asymmetric ({\it skewed}) distribution $P(\delta v_r)$ which can be written in exact closed form in terms of certain special functions, see below. More specifically, we consider
\begin{equation}
P(\delta v_r)= \int_0^\infty P(\delta v_r|\epsilon_\ell)f(\epsilon_\ell) d\epsilon_\ell= \int_0^\infty \frac{1}{\sqrt{2\pi \epsilon_\ell}}\, \exp\left[-{\frac{\left(\delta v_r-\langle \delta v_r|\epsilon_\ell\rangle\right)^2}{2\epsilon_\ell}}\right] f(\epsilon_\ell) d\epsilon_\ell,
%
\label{eq:P1e}
\end{equation}
where the conditional mean of velocity increments $\langle \delta v_r|\epsilon_\ell\rangle$ is a function of $\epsilon_\ell$ with the constraint of null global average, i.e., $\langle \delta v_r\rangle=0$, as required for homogeneous and isotropic turbulence.
We thus make the choice
\begin{align}
\langle \delta v_r | \epsilon_\ell \rangle = \mu  \left({\epsilon_\ell}-{\langle \epsilon_\ell \rangle} \right),
\label{eq:mean}
\end{align} 
where $\mu$ is a flow-related asymmetry parameter 
so that $\langle \delta v_r\rangle=0$ is ensured for any~$\mu$, with the advantage that it renders possible
a closed analytical form for $P(\delta v_r)$. 
We shall see below that the parameter $\mu$ controls the overall asymmetry of the resulting distribution~$P(\delta v_r)$.
%
%
(We can also introduce a dimensionless parameter $b = |\mu|\sqrt{\langle \epsilon_\ell \rangle}$, but for our purposes here it is more convenient to work with $\mu$ itself; see below.)
In Fig.~\ref{fig:gauss} we show qualitatively how a  weighted mixture of Gaussians with nonzero mean (lower curves) can yield an asymmetric, heavy-tailed distribution (uppermost curve). 

The possibility of producing asymmetric distributions  by  compounding Gaussian distributions with nonzero mean as indicated in (\ref{eq:P1e}) has been generally discussed,  e.g., in Refs.~\cite{naert-1998,gagneetal94,Dubrulle2000,valvoetal15}, but with no specific models for the resulting distribution. In Ref.~\cite{castaing_PhysD90} a particular non-Gaussian model was also proposed, although it did not lead to a closed form solution and had the drawback of producing a marginal distribution with nonzero mean. A model for non-Gaussian statistics and intermittency based on an ensemble of Gaussian fields---albeit with zero mean---has been also considered in the literature~\cite{wilczek16}. To the best of our knowledge, the prescription given in Eq.~(\ref{eq:mean}) for the conditional mean velocity $\langle \delta v_r|\epsilon_\ell\rangle$ has not been used before. This is a crucial ingredient that allows us to  obtain an analytic solution for the {\it skewed} marginal distribution~$P(\delta v_r)$.

\begin{figure}[h]
\center \includegraphics[width=0.55\textwidth
]{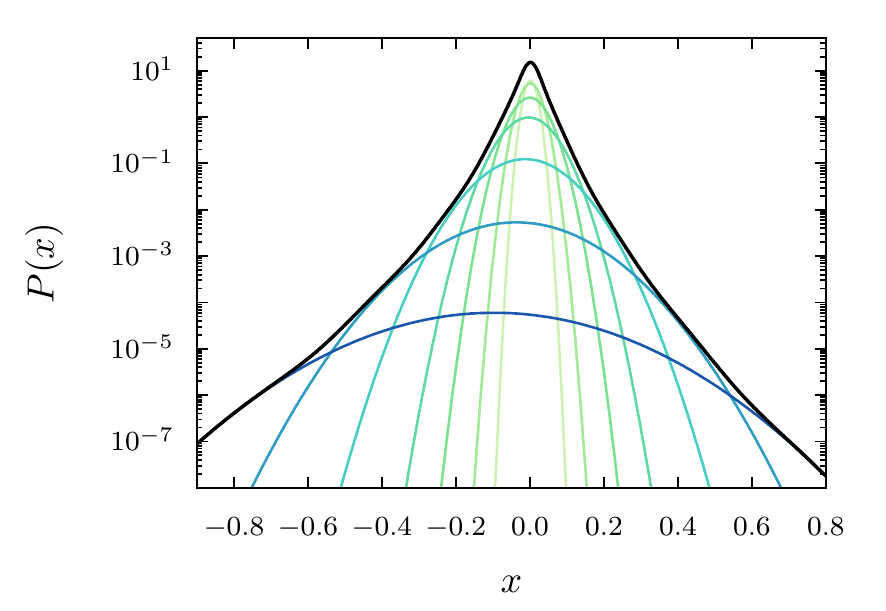}
\caption{Schematic mixture of Gaussians with nonzero mean yielding a skewed heavy-tailed distribution with zero mean. The  uppermost black curve is the sum of the lower curves, which correspond to Gaussians with  variances in the interval [0.0002, 0.05] and means as in Eq.~(\ref{eq:mean}) with $\mu=-2$, multiplied by weights arbitrarily  chosen  for convenience of illustration.}
\label{fig:gauss}
\end{figure}

We now turn to the calculation of the background distribution $f(\epsilon_\ell)$ in Eq.~(\ref{eq:P1e}).
The scale~$\ell$ is assigned to the $N$-th level of the turbulence hierarchy
$(\epsilon_\ell \leftrightarrow \epsilon_N)$,
that is, $\ell=L/2^N$, where $L$~is the integral scale and $N$ is the number of levels in the cascade down from~$L$ to~$\ell$. 
Our hierarchical intermittency model is defined by the following set of~$N$ stochastic differential equations:
\begin{align}
d\epsilon_i =-\gamma _i\left( \epsilon_{i}-\epsilon_{i-1}\right)
\left (1 + \alpha^2 \frac{ \epsilon_{i-1} }{ \epsilon_i } \right ) dt +
\kappa_i \sqrt{ \epsilon_{i}\epsilon_{i-1} } dW_i,
\label{hierarc}
\end{align}
for $i=1,\dots,N$, where $\epsilon_i \ge 0$ represents  the energy transfer rate from the hierarchy level~$i$ to smaller scales,  
$\gamma_i > 0$ is a relaxation rate,  $\kappa_i > 0$ characterizes the strength  of the multiplicative noise (and hence of the intermittency) in the hierarchical level~$i$, and $W_i$~denotes a Wiener process. The intermittency model (\ref{hierarc}) with $\alpha=0$ has been introduced in \cite{pre2017}. The generalization above (with $\alpha\ne0$) is important to consider because
the parameter $\alpha>0$ can be associated with a residual dissipation in the inertial range (see below), which is usually neglected in phenomenological cascade models.

Physically, the deterministic term in Eq.~(\ref{hierarc}) represents the coupling between adjacent scales, whereas
the stochastic term emerges from the complex interactions among all scales and is necessary for intermittency~\cite{pre2017}. %
%
We further observe that a rescaling of variables
$\epsilon_i \to \zeta \epsilon_i$ properly leaves the model dynamics unchanged, which is a required property for a multiplicative cascade model~\cite{jimenez2000} in the sense that it implies $f(\epsilon_i | \epsilon_{i-1})d \epsilon_i= g(x) dx$, for $x=\epsilon_i/\epsilon_{i-1}$, where  $f(\epsilon_i | \epsilon_{i-1})$ is the conditional distribution for $\epsilon_i$ with $\epsilon_{i-1}$ fixed and $g(x)$ is some function of $x$. 
Moreover, one can verify that if $\alpha=0$ then $\langle \epsilon_i\rangle=\epsilon_0$ for $t\to\infty$, whereas for $\alpha\ne 0$ it can be shown 
[see Appendix A, Eq.~(\ref{eq:Amean})]
that $\langle \epsilon_i\rangle/\langle \epsilon_{i-1}\rangle=1-\alpha^2$, as $\alpha \to0$, thus showing that the energy flux leaving the scale~$i$ is actually smaller than that entering it. 
In this sense, it is thus expected that $\alpha$ becomes negligible for very large Reynolds number.
The model above is perhaps the simplest stochastic dynamical model of intermittency that allows for an analytic solution (see below) and incorporates a small degree of dissipation in the cascade, so that it can describe intermittency even at not so high Reynolds numbers where residual dissipation  might be relevant. It is interesting to notice that the nonlinear  relaxation term in (\ref{hierarc}) is similar to the anomalous drift coefficient discussed in~\cite{Nature_Lutz} to model  friction in the context of the unusual transport of cold atoms in dissipative optical lattices. (Higher-order terms could in principle be added in~(\ref{hierarc}) but they should not affect our findings significantly and, besides, destroy the exact solvability of the model.
Other non-exactly-solvable  stochastic models of intermittency were considered, e.g.,~in~\cite{eggers_1992,japanese2003}.)

We  assume that the time scales within the cascade are largely separated, with faster dynamics at smaller scales, 
i.e., $\gamma_N \gg \gamma_{N-1} \gg \cdots \gg \gamma_1$. We  consider furthermore that $\kappa_N \gg \kappa_{N-1} \gg \cdots \gg \kappa_1$, which is reasonable since one expects stronger intermittency at smaller scales, in such a way  that the dimensionless ratio $\beta \equiv 2 \gamma_i / \kappa_i^2$ remains invariant across scales. Under these assumptions,  the stationary solution of the Fokker-Planck equation associated with~(\ref{hierarc}) under It\^o prescription is given by
\begin{equation}
f(\epsilon_i | \epsilon_{i-1}) = \frac{ (\epsilon_i / \epsilon_{i-1} )^{p - 1} }
{ 2 \epsilon_{i-1} \alpha^{p} K_{p} (\omega)}
\exp \left ( -\frac{\beta \epsilon_i}{\epsilon_{i-1}} - \frac{\beta \alpha^2 \epsilon_{i-1}}{\epsilon_{i}} \right ),
\label{gig}
\end{equation}
where $p = \beta (1-\alpha^2)$, $\omega= 2  \alpha \beta$ 
and $K_p (x)$ is the modified Bessel function of second kind.
We notice that the density function~(\ref{gig})
has the form of a generalized inverse Gaussian (GIG) distribution,
which has been applied to describe diverse fluctuation phenomena~\cite{gig1}. 

By denoting
$f(\epsilon_N) \equiv f(\epsilon_\ell)$ in Eq.~(\ref{eq:Px}), 
we write
\begin{equation}
f(\epsilon_N) = \int_0^\infty ... \int_0^\infty f(\epsilon_{N} | \epsilon_{N-1})
\prod_{i=1}^{N-1} [ f(\epsilon_{i} | \epsilon_{i-1}) d \epsilon_{i} ].
\label{compos}
\end{equation}
Notably, these integrals can be performed exactly 
to give 
%
\begin{equation}
f(\epsilon_N) = \frac{1}{\epsilon_0 \left[\alpha K_{p}( \omega)\right]^N}
R_{0, N}^{N, 0}  \left(
\begin{array}{c}
{ - } \\
{ (\boldsymbol p - \boldsymbol 1, \boldsymbol \omega/2) }
\end{array}
\bigg | {\beta}^N \frac{\epsilon_N}{\epsilon_0} \right),
\label{rn}
\end{equation}
where
${\boldsymbol p} \equiv (p, ..., p)$, 
${\boldsymbol \omega} \equiv (\omega, ..., \omega)$,  and $R_{p, q}^{m, n}$ is a new  special function
defined in 
Appendix~A. 
The function
$R_{p, q}^{m, n}$ can be viewed as a generalization of the  
Meijer $G$-function $G_{p, q}^{m, n}$,
in which the gamma functions $\Gamma (\nu)$ are essentially
replaced by the Bessel functions $K_\nu (x)$ in the Mellin transform~\cite{mathai}.

Finally, substituting Eq.~(\ref{rn}) into~Eq.~(\ref{eq:P1e}) and using some properties of the $R$-functions (see Appendices A and B), 
we obtain 
\begin{equation}
P_N(\delta v_r) = c e^{\mu y}
%
R_{0, N+1}^{N+1, 0}  \left(
\begin{array}{c}
{ - } \\
{ [(0, \boldsymbol p - \frac{\boldsymbol 1}{\boldsymbol 2}), [(\frac{|\mu y|}{2}, \frac{ \boldsymbol \omega }{ 2 })] }
\end{array}
\bigg | \frac{{\beta}^N y^2}{2\epsilon_0} \right),
\label{pv}
\end{equation}
with $y = \delta v_r 
+ \mu \langle\epsilon_N\rangle$ and $c = ( 2/\pi \epsilon_0 \alpha^N )^{1/2}/[K_{p}( \omega)]^N$. 
For a given $N$, the distribution above has four  parameters, namely: $\alpha$, $\beta$, $\epsilon_0$, and $\mu$. 
As discussed above, the parameter~$\alpha$ is physically related to a residual energy dissipation in the inertial range. 
On the other hand, the dimensionless constant~$\beta$ together with~$\epsilon_0$ define a typical scale $(2 \epsilon_0 / \beta^N)^{1/2}$ for the fluctuations of the velocity increments~$\delta v_r$,
so that a larger relative noise (intermittency) strength and/or a lower relaxation rate consistently yields a broader distribution~$P_N(\delta v_r)$. Lastly, the parameter $\mu$ controls the asymmetry of the distribution, as already mentioned. 

At this point, we emphasize that, although our model has four free parameters, they are determined in pairs---first~$\epsilon_0$ and~$\mu$, then~$\alpha$ and~$\beta$---in a two-step procedure involving the background distribution  $f(\epsilon_N)$, which is a more stringent constraint than a direct fit of $P_N(\delta v_r)$; see Section~\ref{sec:data}. 
Indeed, the fact that the background distribution~$f(\epsilon_N)$ is available to fit the empirical data in an unambiguous way, as seen below, actually proves to be an important feature of our method, since it is known that the distribution of velocity increments~$P_N(\delta v_r)$ can be almost equally well fitted by different theoretical expressions, thus making it difficult to  select between competing models~\cite{pre2017}.

We note that the single-scale case, i.e.,
$N = 1$, in Eq.~(\ref{pv}) corresponds to the generalized hyperbolic
distribution, as the distribution $P_1(\delta v_r)$ in this case reduces to  a Gaussian variance-mean mixture where the mixing distribution is the GIG distribution; see Eqs.~(\ref{eq:Px})-(\ref{eq:mean}) and~(\ref{gig}). 
The  generalized hyperbolic
distribution has found many applications, including  in the analysis of turbulent velocity increments \cite{NIG}.
It appears, however, that the $N > 1$ multiscale scenario and the corresponding $R$-distribution defined in (\ref{pv}) have not been considered before in the literature. We anticipate here that the multiscale behavior ($N>1$) is crucial to generate heavy tails, as the case $N=1$ yields only semi-heavy tails; see below.
We also highlight that $P_N(\delta v_r)$ given by Eq.~(\ref{pv}) is negatively (positively) skewed for
$\mu < 0$   ($\mu > 0$), whereas for $\mu = 0$ a symmetric (non-skewed) distribution arises. 

The large-$|\delta v_r|$ behavior of $P_N(\delta v_r)$ evidences
the presence of non-Gaussian tails.
Indeed, for $N > 1$ and negative asymmetry, $\mu<0$, we obtain
\begin{align}
P_N(\delta v_r) 
\sim |y|^{\theta}  \exp\left[-{\beta}N\left(\frac{y}{\epsilon_0 |\mu|} \right)^{1/N} \right] g(\delta v_r),
\label{eq:sexp}
\end{align}
where $\theta=  p  + 1/(2N) - 3/2$ and  $g(\delta v_r) = 1$ for $\delta v_r \to -\infty$ and $g(\delta v_r) = e^{-2|\mu |y}$ for $\delta v_r \to +\infty$.
The negatively skewed marginal distribution displays an asymptotic behavior to the right $(\delta v_r \to +\infty)$ 
with exponential decay,
while the left tail is heavier, in the form of a modified  stretched exponential. 
In contrast, for $N=1$ modified exponential tails emerge on both sides: $P_{N=1}(\delta v_r) \sim z^{p -1}e^{\mu y-\kappa z}$, 
where $\kappa=\sqrt{\mu^2+2\beta/\epsilon_0}$ and  $z=\sqrt{y^2+2\alpha^2\beta\epsilon_0}$ for~$\delta v_r \to\pm\infty$.
Stretched exponentials have for long been used to fit turbulence data~\cite{Kailetal92} despite the lack of a theoretical basis for this. Our model thus provides a reasonable physical framework for the emergence of such heavy tailed distributions. 

\section{Data analysis} 
\label{sec:data}

We now describe how to apply the above formalism to the data analysis of turbulent flows.

Consider a large dataset $\{ \delta v_r (j) \}$ of longitudinal velocity increments, with $j = 1, ..., N_v$. As a first step, we need to determine the optimal window size $M$ over which the variance of $\delta v_r (j)$ is supposed to remain approximately constant.
By dividing the original series into overlapping intervals of size~$M$, we define~\cite{pre2017,nature2017}
an estimator of the local variance for each interval as
$\epsilon (k) = \sum_{j=1}^{M} [\delta v (k-j) - \overline{\delta v}(k)]^2/M$,
where $\overline{\delta v}(k) = \sum_{j=1}^{M}\delta v (k-j) /M$, with $k = M, ..., N_v$. 
As discussed in Section~\ref{sec:theory}, we take the variance of $\delta v_r$ over a region of size $\ell=Mr$ as a proxy measure for the energy transfer rate from scale $\ell$ to smaller scales~\cite{naert-1998,gagneetal94}. 
For various choices of~$M$ and varying the asymmetry parameter~$\mu$ for each~$M$, we numerically compound the empirical distribution of the variance series $\{ \epsilon(k) \}$ with the Gaussian as given in~(\ref{eq:P1e}), and select the optimal parameters~$M$ and~$\mu$ for which the compounding integral (\ref{eq:Px}) best fits the distribution of velocity increments computed from the original data. 
(See, e.g.,~\cite{guhr-beck-swinney,guhr-beck-swinney2,guhr-beck-swinney3} for other methods to estimate the optimal window size for the variance series in the case of superposition of Gaussians with {\it zero} mean, therefore not corresponding to our context.) 
The knowledge of $M$ then allows to express~$\epsilon_0$ in terms of the mean~$\langle\epsilon_N\rangle$ of the variance series and the parameters~$\alpha$ and~$\beta$ (see Appendix C), thus leaving only $\alpha$ and $\beta$ to be determined.

\begin{figure}[t]
\begin{center}
{\includegraphics[width=0.45\textwidth]{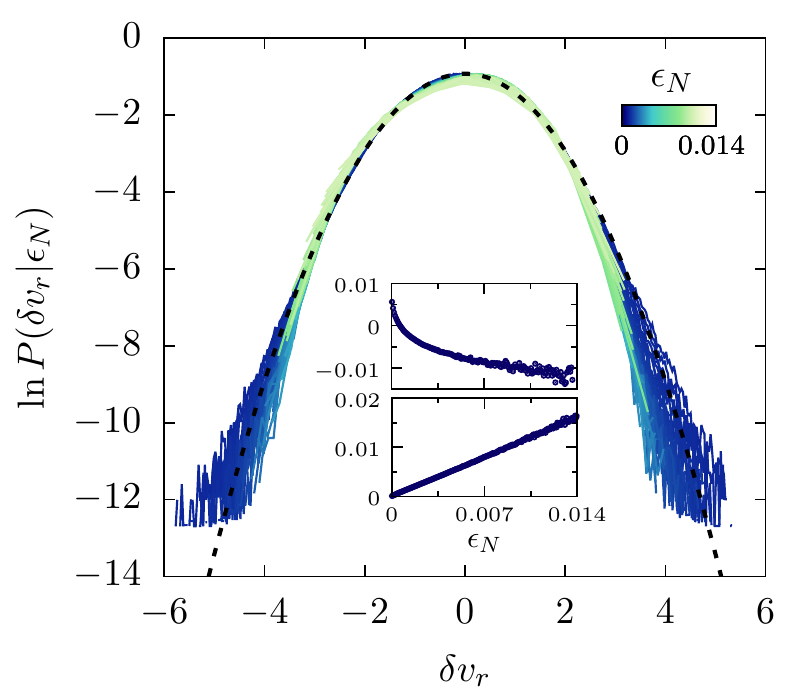}}
\end{center}
\caption{Conditional distribution of velocity increments for DNS turbulence data
of a system of size $1024^3$ and Reynolds number $\mbox{Re}_\lambda \approx 433$~\cite{jhu08}.
Distinct distributions obtained for the optimal window size $M=19$ have been rescaled and shifted to have the same mean~(zero) and variance~(unity). A nice agreement is observed with a Gaussian of zero mean and unity variance (dashed line). }
\label{fig:cond}
\end{figure}

Once $M$ is set, we estimate the number $N$ of scales in the cascade by $N=\log_2(L/\ell) = \log_2(L/Mr)$, where $L$ is the integral scale; 
see discussion preceding Eq.~(\ref{hierarc}).
After obtaining the variance series $\{\epsilon_{N}(k)\}$, we fit the empirical distribution $f(\epsilon_N)$
to  Eq.~(\ref{rn}) to determine $\alpha$~and~$\beta$. 
Finally, the theoretical distribution of velocity increments $P_N(\delta v_r)$ is computed by inserting the parameters in~(\ref{pv}).
Therefore, we remark that the setting of parameters is completed prior to the calculation of~$P_N(\delta v_r)$.

Let us now apply this procedure to the analysis of isotropic turbulence data~\cite{jhu08} generated 
by the extensive direct numerical simulation (DNS) of the Navier-Stokes equations  for a system with $1024^3$ lattice points in a periodic cube and Taylor-based Reynolds number $\mbox{Re}_\lambda \approx 433$.
The dataset was obtained from the Johns Hopkins University turbulence research group's database~\cite{jhu08}.
The simulation 
spans five large eddy turnover times, 
from which we considered $\approx 3 \times 10^8$ points for our statistics. To test our intermittency model and show that it applies well to turbulence data, we shall analyze here  the velocity increments $\delta v_r$ computed at the smallest resolved scale~$r$, which lies in the near dissipation range as $r \approx 2.14\eta$~\cite{jhu08}, where~$\eta$ denotes the Kolmogorov scale. A more complete analysis including other scales~$r$ will be left for future studies.

We begin by analyzing the conditional distribution $P(\delta v_r|\epsilon_\ell)$, which requires computing first  the joint distribution $P(\delta v_r,\epsilon_\ell)$~\cite{castaing_PhysD90,gagneetal94,naert-1998,homannetal11}. To this end, we  adopt here the following  ad-hoc prescription: for each window of size $M$ of the dataset $\{ \delta v_r (j) \}$ we compute the corresponding variance $\epsilon_{N}(k)$ and associate it with the velocity increment $\delta v_r$ at the center of the  respective window. The variance series $\{\epsilon_{N}(k)\}$ thus generated is then `binarized' and for each bin we compute the respective histogram $P(\delta v_r|\epsilon_N)$ of  velocity increments.
In Fig.~\ref{fig:cond} we  show that the empirical conditional distributions $P(\delta v_r|\epsilon_N)$ obtained for $M=19$  are indeed well described  by Gaussians, thus validating the assumption (\ref{eq:P1e}), 
with the upper (lower) inset  displaying the conditional mean (variance).
Although the  observed behavior of $\langle \delta v_r | \epsilon_N\rangle$ vs.~$\epsilon_N$ is only approximately linear, 
the important point to note is that $\langle \delta v_r | \epsilon_N \rangle$ decreases from a positive value  to a negative one as $\epsilon_N$ increases, thus implying $\mu<0$.
(Models with a nonlinear mean-variance relationship could in principle be introduced but  the distributions may not be given in analytical form.)
%
%
A similar trend as that seen in the upper inset of Fig.~\ref{fig:cond} has been observed before, e.g., in Refs.~\cite{hosokawa1994, gagneetal94, naert-1998}, although there the velocity increments $\delta v_r$ and the variance $\epsilon_r$ are computed over the same scale $r$, while in our case $\epsilon_\ell$ is defined over a larger scale $\ell=Mr$; see the discussion after Eq.~(\ref{eq:Px}).

\begin{figure}[t!]
\centering
{\includegraphics[width=0.95\textwidth]{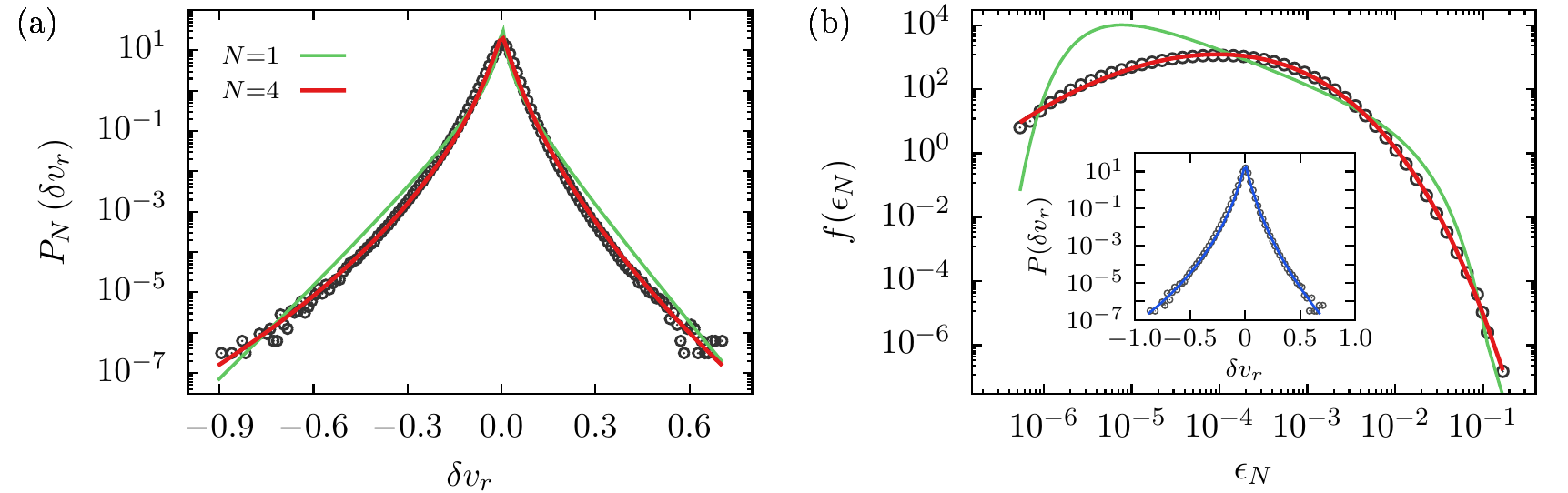}}
\caption{(a) Distribution of velocity increments and (b) background density
of local variances for the  DNS data of Fig.~\ref{fig:cond} (circles).
Excellent fits to the theoretical results (red lines), Eqs.~(\ref{rn}) and~(\ref{pv}), respectively, are shown for $N=4$ scales.
For comparison, the case  
with $N = 1$ (single scale) is also plotted (green lines),  displaying much poorer fits. 
Inset in (b): nice agreement of the empirical distribution of velocity increments (circles) and the compounding integral (blue line), Eq.~(\ref{eq:P1e}), of the Gaussian and the density $f(\epsilon_N)$ obtained from the DNS data.}
\label{fig:exp}
\end{figure}

Our analysis goes further, however, in that it shows mathematically that such `local behavior' of the average velocity increment is linked to both the `global asymmetry' and the non-Gaussian tails of the marginal distribution of the velocity increments; see the role of~$\mu$ in Eqs.~(\ref{pv}) and~(\ref{eq:sexp}).
Physically, the change in 
$\langle \delta v_r | \epsilon \rangle$ from positive to negative values for increasing energy dissipation rate $\epsilon$,  as inferred from Fig.~\ref{fig:cond}, is a clear  evidence that the emergence of skewness  is directly related to  intermittency: in regions of small (large) $\epsilon$ the fluid particle is more likely to accelerate (decelerate) from one point to the next, resulting in a positive (negative) local average  $\langle \delta v_r | \epsilon \rangle$, so that the long-time statistics of~$\delta v_r$ 
has zero mean but negative skewness.
We remark that a link between intermittency and skewness governed by a single parameter was also found in a recent stochastic model for the turbulent velocity field \cite{JFM16}, but no explicit distribution is obtained there.

We now proceed to further test the model.
By applying the  optimization procedure  described above to select $M$ and $\mu$  we obtain $\mu = -1.82$ and $M=19$ (yielding $\langle\epsilon_N\rangle = 1.09 \times 10^{-3}$), 
which leads to the nice agreement in the inset of Fig.~\ref{fig:exp}(b) between the numerical compounding (solid line) of the Gaussian with the {\it empirical} $f(\epsilon_N)$, see Eq.~(\ref{eq:P1e}), and the velocity increments distribution from the DNS 
data (circles). 
We note furthermore that the scale~$\ell=Mr$ belongs to the inertial range, since $\ell \approx 40.7\eta$ nearly coincides with the Taylor scale $\lambda \approx 41.1\eta$~\cite{jhu08}, thus confirming the separation of scales anticipated in the discussion of Eq.~(\ref{eq:Px}).
Using that  the integral scale in this case  \cite{jhu08} is $L=104.7\eta=224 r$, 
we estimate the number of scales in the model hierarchy: $N= \log_2(L/\ell)=\log_2(224r/19r)\approx 4$.

\begin{figure}[t!]
\centering
{\includegraphics[width=0.57\textwidth]{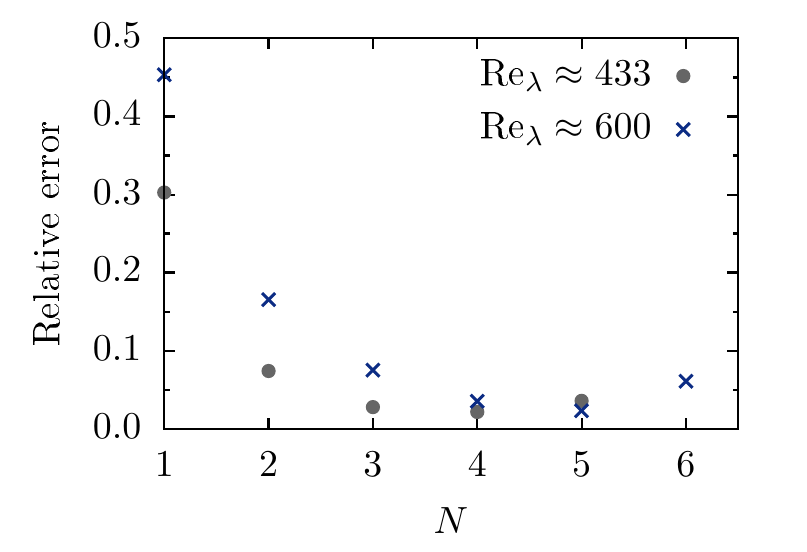}}
\caption{Relative error for the background distributions $f(\epsilon_N$) obtained from fits with different hierarchy levels $N$. The optimal values of $N$ for both DNS datasets, i.e., $N=4$ for $\mbox{Re}_\lambda\approx 433$ (solid circles) and $N=5$ for  $\mbox{Re}_\lambda\approx 600$ (crosses), correspond to the estimates provided by comparing the respective integral length scales to the scale $\ell=Mr$ over which $\epsilon_N$ is computed (see text).}
\label{fig:errors}
\end{figure}

Figure~\ref{fig:exp}(a) and the main panel of Fig.~\ref{fig:exp}(b) display, respectively,  
the marginal distribution~$P_N(\delta v_r)$ and background density $f(\epsilon_N)$ for $N=4$. 
The theoretical results are shown in solid lines and the empirical data are depicted in circles, 
with excellent agreement observed in both $P_N(\delta v_r)$ and $f(\epsilon_N)$. 
The best fit parameters are $\alpha = 0.17$ and $\beta = 2.72$.  
For comparison, we also plot the best fit using $N=1$ (single scale), which clearly does not perform
so well as the one with $N=4$. 
This result evidences that this DNS dataset cannot be properly described  with only a single scale.
We also confirmed that the $N=4$ case indeed produces a better fit than $N = 2, 3, 5$. This is depicted in Fig.~\ref{fig:errors}, in which we show the relative squared error (solid circles)  of the fitted background  distribution $f(\epsilon_N)$ 
for different hierarchy levels~$N$.

We stress that {\it no} curve fitting was performed in Fig.~\ref{fig:exp}(a); the fit was done only in the main panel of  Fig.~\ref{fig:exp}(b) to obtain the parameters $\alpha$ and $\beta$ entering the density $f(\epsilon_N)$. Once these parameters were known, we simply plotted $P_N(\delta v_r)$ using Eq.~(\ref{pv}) and superimposed it with the empirical histogram for the velocity increments $\delta v_r$.
Thus, the nice agreement exhibited for $N=4$ in Fig.~\ref{fig:exp}(a) using the  parameters determined from Fig.~\ref{fig:exp}(b) attests  to the method's self-consistency.

\begin{figure}[t]
\centering
{
\includegraphics[width=0.98\textwidth]{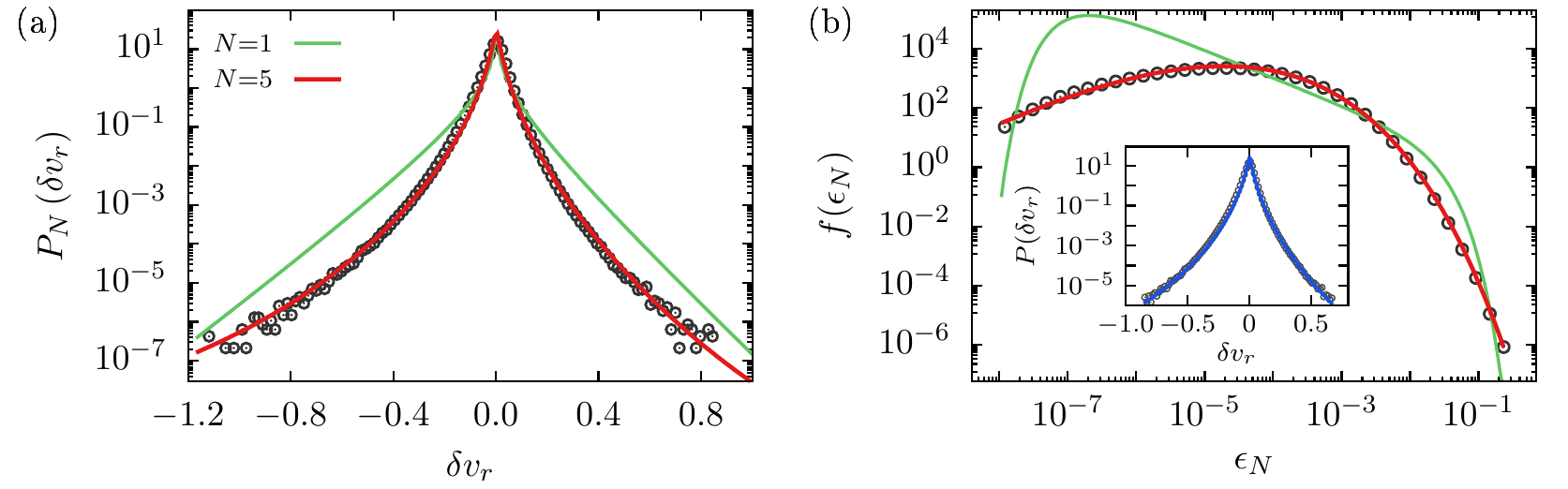}}

\caption{(a) Distribution of velocity increments and (b) background density  for a  system with 4096$^3$ points and  Reynolds number $\mbox{Re}_\lambda \approx 600$~\cite{4096}.
The nice fit of the DNS data (circles) to the theoretical model (red lines) occurs for $N=5$ scales.
A poor fit is noticed in green lines for $N=1$.
Inset: description as in the inset of Fig.~\ref{fig:exp}(b).}
\label{fig:dns}
\end{figure}

We now turn to analyze more recent turbulence data~\cite{4096} from the DNS of the Navier-Stokes equations for a larger system with 4096$^3$ points and higher $\mbox{Re}_\lambda \approx 600$. The dataset consists of only one snapshot in time from which we took~$\approx 3\times 10^8$ points, with the smallest resolved scale being $r \approx 1.11\eta$ and the  integral scale $L=907r$~\cite{4096}.   
(Here again, the analysis of other scales lies out of the scope of this work.)

The theoretical results (solid lines) and DNS data (circles) for $P_N(\delta v_r)$ (main panel) and  $f(\epsilon_N)$ (inset) are shown in Fig.~\ref{fig:dns}.
Here we find $\mu = -1.50$ and $M=27$ (yielding $\langle \epsilon_N \rangle = 9.06 \times 10^{-4}$), 
whereas $\alpha \approx 0$ and $\beta = 2.55$.
From the data in~\cite{4096}  we  obtain a larger number of scales $N=\log_2(907r/27r) \approx 5$.  
Indeed, for $N=5$ a remarkable agreement with the empirical data is observed for both $P_N(\delta v_r)$ and $f(\epsilon_N)$, as seen in Fig.~\ref{fig:dns}(a) and the the main panel of Fig.~\ref{fig:dns}(b), respectively (red curves). As in the previous analysis, the  fit (green curve) using only a single scale $(N=1)$  is not as accurate as that with $N=5$. 
Accordingly, we found that the cases $N = 2, 3, 4, 6$ also led to poorer fits when compared to $N = 5$ (see Fig.~\ref{fig:errors}).

Let us now briefly examine the behavior of the model parameters with Reynolds number. First, note that the larger $N$ obtained for the second dataset, which has a higher $\mbox{Re}_\lambda$, is consistent with the fact that  $L/\eta$ increases with $\mbox{Re}_\lambda$, and so we expect more steps  in the cascade (hence a  larger $N$) as $\mbox{Re}_\lambda$ enhances. 
Furthermore, the fact that $\alpha \approx 0$ for $N=5$ also agrees with the suggested interpretation that~the $\alpha$-term in~Eq.~(\ref{hierarc}) represents a residual dissipation in the inertial range, which is expected to become negligible for very large~$\mbox{Re}_\lambda$, as commented above. Note also that the asymmetry parameter $\mu$   is smaller in magnitude for the second dataset, as expected,  since this case corresponds to higher $\mbox{Re}_\lambda$   and smaller $r$. (Recall that in the second dataset $r/L$ decreases by a factor of four and $r/\eta$, by a factor of 2.)

Moreover, the parameter $\beta$, which controls the shape of the background distribution $f(\epsilon_\ell)$, was found to decrease slightly in the second dataset, implying that $f(\epsilon_\ell)$ is broader in this case; compare the main plots of Figs.~\ref{fig:exp}(b) and~\ref{fig:dns}(b). This behavior is consistent  with the expected ``amplification  of intermittency''~\cite{chevetal05}  as $r$ decreases. If this trend persists for higher $\mbox{Re}_\lambda$ and smaller $r$, then the normalized moments of the velocity derivative distribution should diverge  for $\mbox{Re}_\lambda\to\infty$ (see below).  At present, however,  one cannot rule out the possibility that $\beta$ eventually becomes an increasing function of $N$  as $r$ gets smaller and $\mbox{Re}_\lambda$ larger, which  would lead to constant normalized moments. It thus follows from this discussion that how $\beta$ varies with $N$ is crucial to determine the statistical properties of  velocity increments at small scales. To see this more explicitly, we recall that for $\alpha\to0$ the intermittency model given in Eq.~(\ref{eq:vareps}) recovers that described in Ref.~\cite{pre2017} for which the normalized moments of the background variable $\epsilon_\ell$ are %
\begin{equation}\label{eq:eq}
\frac{\langle \epsilon_\ell^q\rangle}{\epsilon_0^q} = \prod_{j=1}^{q-1}\left(1+\frac{j}{\beta}\right)^N.
\end{equation}
If we assume that our hierarchical model remains valid at very small scales,  it follows from (\ref{eq:P1e}) [in the limit that $\mu\to0$] that
\begin{equation}\label{eq:Sq}
S_{2q}=\frac{\langle (\partial_x v)^{2q}\rangle}{\langle \partial_x v^2\rangle^{q}}\sim \frac{\langle \epsilon_\ell^q\rangle}{\epsilon_0^q}.
\end{equation}
One then sees from Eqs.~(\ref{eq:eq}) and (\ref{eq:Sq}) that  as $N\to \infty$ two quite distinct scenarios arise: if $N/\beta\to\infty$ then the normalized derivative moments $S_{2q}$ diverge, as in Kolmogorov's 1962 theory (K62) \cite{Kol62}, whereas if $N/\beta\to0$ then $S_{2q}= \mbox{constant}$, as predicted by Kolmogorov's original 1941 theory (K41) \cite{K41a,K41b}.

In this context, it is interesting to point out that it has recently been suggested \cite{PRF2019} that one should  observe an approach 
towards the predictions of K41 (rather than K62) as $\mbox{Re}_\lambda$ continues to increase.
Our hierarchical theory described above thus suggests an alternative way to assess this claim 
through  a careful analysis of the behavior of $\beta$ for increasing  $\mbox{Re}_\lambda$. If K41 is indeed to be obtained in such limit one should observe a faster growth of $\beta$ in comparison to $N$,  whereas if $\beta$ continues to decrease (or eventually increases but slower than $N$) then $K62$ is favored.
This interesting possibility certainly deserves further investigation.

In summary, we have seen from the preceding discussion that our model is rather versatile in that the changing behavior of the distribution of velocity increments with varying~$r$ can be well accommodated in the background distribution~$f(\epsilon_N)$.
Further investigation of the model at  more scales, as well as of  the dependence of the model parameters with the Reynolds number, will be left for future studies.

\section{Conclusions}

We have developed a hierarchical model to investigate the emergence of the negative skewness and non-Gaussian behavior of the distribution of short-scale velocity increments in isotropic turbulence. 
The fine agreement between the theoretical  distributions, given in terms of a newly-defined transcendental function,
and the empirical histograms from two independent numerical datasets highlights the crucial role of the multiple scales of the intermittent energy cascade. 

The general character and plasticity of our formalism make it readily adaptable to investigate  
the  emergence of skewed (non-Gaussian) statistics in other complex systems. For example, the symmetric version of our theory has been successfully applied to explain the emergence of turbulence in a photonic random laser, as recently reported in~\cite{nature2017}; and so we expect that the asymmetric model introduced here should also have great applicability.
Indeed, the compounding of an asymmetric conditional Gaussian distribution with a background density built from a hierarchical stochastic model  might be a common feature in contexts as diverse as financial markets~\cite{complex1} 
and biological systems~\cite{complex2}. 
%

In conclusion, we emphasize that the  formalism presented here not only advances on the statistical description of turbulent phenomena but can also be applied to investigate the origin of skewed non-Gaussian distributions in other complex systems.

\begin{acknowledgments}
 This work was supported in part by the following Brazilian agencies: Conselho Nacional de Desenvolvimento Cient\'ifico e Tecnol\'ogico (CNPq), under Grants No.~303772/2017-4, No.~305062/2017-4 and No. 311497/2015-2, Coordena\c c\~ao de Aperfei\c coamento de Pessoal de N\'ivel Superior (CAPES), and Funda\c{c}\~ao de Amparo \`a Ci\^encia e Tecnologia do Estado de Pernambuco (FACEPE), under Grants No.~APQ-0073-1.05/15 and No.~APQ-0602-1.05/14. 
\end{acknowledgments}

\appendix

\section{Derivation of the background distribution}

We start by providing additional details on the exact calculation of the background probability density, $f(\epsilon_N) \equiv f(\epsilon_\ell)$, 
which incorporates the crucial effect of the fluctuating energy flux (intermittency) on the turbulence properties.
As indicated in Eq.~(\ref{eq:Px}), 
in this scenario the marginal distribution of short-scale velocity increments, $P_N(\delta v_r) \equiv P(\delta v_r)$, 
is obtained by compounding $f(\epsilon_\ell)$ with 
the Gaussian conditional distribution $P(\delta v_r|\epsilon_\ell)$.

Our starting point is the multiple integral representation of the background density, Eq.~(\ref{compos}),
\begin{equation}
f(\epsilon_N) = \displaystyle \int\limits_{0}^{\infty} ... \displaystyle \int\limits_{0}^{\infty} f(\epsilon_{N} | \epsilon_{N-1})
\prod_{i=1}^{N-1} [ f(\epsilon_{i} | \epsilon_{i-1}) d \epsilon_{i} ],
\label{sm:compos}
\end{equation}
in which the generalized inverse Gaussian (GIG) distribution, 
\begin{equation}
f(\epsilon_i | \epsilon_{i-1}) = \frac{ (\epsilon_i / \epsilon_{i-1} )^{p - 1} }
{ 2 \epsilon_{i-1} \alpha^{p} K_{p} (\omega)}
\exp \left ( -\frac{\beta \epsilon_i}{\epsilon_{i-1}} - \frac{\beta \alpha^2 \epsilon_{i-1}}{\epsilon_{i}} \right ), 
\label{sm:gig}
\end{equation}
arises as the solution of the system of stochastic differential equations, Eq.~(\ref{hierarc}),  
with $\kappa_i = \sqrt{{2\gamma_i}/{\beta}}$, $p = \beta (1-\alpha^2)$, $\omega = 2\alpha\beta$,
and $K_\nu (x)$ as the modified Bessel function of second kind. Introducing the new variable $x_i=\epsilon_i/\epsilon_{i-1}$, we write
\begin{equation}
f(\epsilon_i|\epsilon_{i-1})\, d\epsilon_i = g_i(x_i)\, dx_i,  
\label{NF:15}
\end{equation}
where 
\begin{equation}
g_i(x_i)= \frac{x_i^{p-1}}{2\alpha^{p} K_{p} (\omega)}  \exp \left ( -\beta x_i - \frac{\beta \alpha^2}{x_i} \right ).
\label{NF:12:01}
\end{equation}
We proceed by observing that 
	\begin{equation}
	\begin{aligned}
	\epsilon_N &=  \frac{\epsilon_N}{\epsilon_{N-1}} \frac{\epsilon_{N-1}}{\epsilon_{N-2}} \dots \frac{\epsilon_1}{\epsilon_{0}} \epsilon_0  \\
	&=  \epsilon_0\,  \prod_{j=1}^{N} x_j.
	\label{NF:17}
	\end{aligned}
	\end{equation}

Next we recall that the Mellin transform~\cite{mellin} of a function $f(x)$ is defined by 
	\begin{equation}
	\begin{aligned}
	\widetilde{f}(s) & \equiv   \displaystyle \int\limits_{0}^{\infty}  \varepsilon^{s-1}    f (x)       \,dx ,
	\label{eq:ftil}
		\end{aligned}
	\end{equation}
which implies the following relation between the Mellin transforms of $f(\epsilon_N)$ and  $g_i(x)$:
	\begin{equation}
	\begin{aligned}
	\widetilde{f}(s) & \equiv   \displaystyle \int\limits_{0}^{\infty}  \epsilon^{s-1}_N    f(\epsilon_N)       \,d\epsilon_{N}   \\
    &= \displaystyle \int\limits_{0}^{\infty} ... \displaystyle \int\limits_{0}^{\infty} \epsilon^{s-1}_N     \prod_{i=1}^{N}  [ f(\epsilon_i |\epsilon_{i-1}) ]           \,d\epsilon_{N} \dots d\epsilon_1   \\
	&= \epsilon^{s-1}_0 \prod_{i=1}^{N} \left[ \displaystyle \int\limits_{0}^{\infty}  x^{s-1}_i      g_i(x_i)dx_i
	\right]\\
	&=\epsilon^{s-1}_0 \prod_{i=1}^{N} \widetilde{g}_i(s)    .
	\label{NF:18}
	\end{aligned}
	\end{equation}
%
%

The Mellin transform of Eq.~(\ref{NF:12:01}) is~\cite{mellin}
	\begin{equation}
	\widetilde{g}_i(s) =  \alpha^{s-1}\frac{K_{s+p-1}(\omega)}{K_{p}(\omega)}.
	\label{NF:20}
	\end{equation}
Inserting Eq.~(\ref{NF:20}) into Eq.~(\ref{NF:18}), we see that the Mellin transform of $f(\epsilon_N)$ is
	\begin{equation}
	\begin{aligned}
	\widetilde{f}(s) & =    \epsilon^{s-1}_0     
\left [	\alpha^{s-1}\frac{K_{s+p-1}(\omega)}{K_{p}(\omega)} \right ]^N
	\\
	&=   \left(\epsilon_0\,\alpha^N\right)^{s-1}  
\left [    \frac{K_{s+p-1}(\omega)}{K_{p}(\omega)} \right ]^N.
	\label{NF:21}
	\end{aligned}
	\end{equation}
%
Now, using Eq.~(\ref{NF:21}) and the formula of the inverse Mellin transform, we can write $f(\epsilon_N)$ as the contour integral
	\begin{equation}
	f(\epsilon_N) =  \frac{1}{\epsilon_0 \left [\alpha K_{p}(\omega) \right]^N } \frac{1}{2\pi i}     \displaystyle \int\limits_{\Gamma}  \left(\frac{\epsilon_N }{\epsilon_0 \alpha^N }\right)^{-s}       
    \left [ K_{s+p-1}(\omega) \right ]^N  ds.
	\label{NF:24}
	\end{equation}

Further progress can be made by introducing a generalization of the Meijer-$G$ function~\citep{mathai}, which we shall refer to as the $R$-function, in terms of the following Mellin-Barnes integral, 
	\begin{equation}
	R_{p,q}^{m,n} \left(\begin{matrix} \pmb{a},\pmb{A} \\
	\pmb{b},\pmb{B}
	\end{matrix} \bigg| x \right)  = \frac{1}{2\pi i}     \displaystyle \int\limits_{\Gamma}   x^{-s}  \widetilde{R}_{p,q}^{m,n} \left(\begin{matrix} \pmb{a},\pmb{A} \\
	\pmb{b},\pmb{B}
	\end{matrix} \bigg| s\right)     ds,
	\label{NF:31}
	\end{equation}
	where
	\begin{equation}
	\widetilde{R}_{p,q}^{m,n} \left(\begin{matrix} \pmb{a},\pmb{A} \\
	\pmb{b},\pmb{B}
	\end{matrix} \bigg| s\right)   = \frac{ \displaystyle \prod_{j=1}^{m}    B^s_j  K_{{b_j}+s}(2B_j) \displaystyle \prod_{k=1}^{n}    A^{-s}_k  K_{1-a_k-s}(2A_k) }{ \displaystyle \prod_{k=n+1}^{p}    A^s_k  K_{a_k+s}(2A_k)  \displaystyle \prod_{j=m+1}^{q}    B^{-s}_j  K_{1-b_j-s}(2B_j)},
	\label{NF:32}
	\end{equation}
	and $\pmb{a} = (a_1,\allowbreak\dots,\allowbreak a_n,\allowbreak a_{n+1},\allowbreak\dots,\allowbreak a_p)$, $\pmb{A} = (A_1,\allowbreak\dots,\allowbreak A_n,\allowbreak A_{n+1},\allowbreak\dots,\allowbreak A_p)$, $\pmb{b} = (b_1,\allowbreak\dots,\allowbreak b_m,\allowbreak b_{m+1},\allowbreak\dots,\allowbreak b_q)$,  $\pmb{B} = (B_1,\allowbreak\dots,\allowbreak B_m,\allowbreak B_{m+1},\allowbreak\dots,\allowbreak B_q)$.
The contour path $\Gamma$ is chosen 
so that the conditions for the existence of the inverse Mellin transform are satisfied~\cite{mellin},  since $R$ and $\widetilde{R}$ are Mellin pairs (see also below). From Eq.~(\ref{NF:32}) we see that
    \begin{equation}
	R_{0,N}^{N,0} \left(\begin{matrix} - \\
	\pmb{b},\pmb{B}
	\end{matrix} \bigg| x \right)  = \frac{1}{2\pi i}     \displaystyle \int\limits_{\Gamma}   x^{-s}       \prod_{j=1}^{N}    B^s_j  K_{{b_j}+s}(2B_j)  ds,
	\label{NF:30}
	\end{equation}
	where $\pmb{b} = (b_1,\dots,b_N)$ and $\pmb{B} = (B_1,\dots,B_N)$. 

The newly defined special function
$R_{p, q}^{m, n}$ can be viewed as a generalization of the Meijer-$G$ function $G_{p, q}^{m, n}$,
in which the gamma functions $\Gamma (\nu)$ are essentially
replaced by $K_\nu (x)$ Bessel functions in the Mellin-Barnes integral above, Eqs.~(\ref{NF:32}) and~(\ref{NF:30}).
Indeed, by using the limit form
$K_\nu (x) \to \Gamma (\nu) 2^{\nu - 1} x^{-\nu}$, $\nu > 0$, $x \to 0$,
we observe that $R_{p, q}^{m, n} \to a \hspace{0.03 cm} G_{p, q}^{m, n}$, 
where $a$ is a constant, 
when the argument of the Bessel functions tends to zero. 

Finally, by comparing Eq.~(\ref{NF:24}) with Eq.~(\ref{NF:30}) we obtain the expression for the background density, Eq.~(\ref{rn}):
\begin{equation}
f(\epsilon_N) = \frac{1}{\epsilon_0 \left [\alpha K_{p}(\omega) \right]^N}
R_{0, N}^{N, 0}  \left(
\begin{array}{c}
{ - } \\
{ (\boldsymbol p - \boldsymbol 1, \boldsymbol \omega/2) }
\end{array}
\bigg | \beta^N \frac{\epsilon_N}{\epsilon_0} \right),
\label{sm:rn}
\end{equation}
where 
${\boldsymbol p}\equiv (p, ..., p)$ and ${\boldsymbol\omega}\equiv (\omega, ..., \omega)$.

The last step consists in compounding  Eq.~(\ref{sm:rn}) with the Gaussian conditional distribution via Eq.~(\ref{eq:P1e}) to obtain exactly the marginal distribution of short-scale velocity increments, $P_N(\delta v_r)$, Eq.~(\ref{pv}), which is also given in terms of an $R$-function. 
To see this, note that the Gaussian distribution in Eq.~(\ref{eq:P1e}) can be written as
\begin{equation}
P(\delta v_r|\epsilon_\ell) = \frac{e^{\mu\,y}}{\sqrt{\pi} } \left( \frac{2\,\mu}{y} \right)^{\frac{1}{2}} R_{1,0}^{0,1}  \left(
\begin{array}{c}
{ \frac{1}{2},\frac{|\mu\, y|} {2} } \\
{ - }
\end{array}
\bigg |  \frac{2\,\epsilon_{\ell}}{y^2} \right),
\label{Eq7:01}
\end{equation}
with $y = \delta v_r 
+ \mu \langle\epsilon_\ell\rangle$. Thus, the statistical composition of Eqs.~(\ref{sm:rn}) and~(\ref{Eq7:01}) is performed using the integral involving the product of two $R$-functions (see property~(\ref{NF:36}) below). We thus find
\begin{equation}
P_N(\delta v_r) = c e^{\mu y}
%
R_{0, N+1}^{N+1, 0}  \left(
\begin{array}{c}
{ - } \\
{ [(0, \boldsymbol p - \frac{\boldsymbol 1}{\boldsymbol 2}), [(\frac{|\mu y|}{2}, \frac{ \boldsymbol \omega }{ 2 })] }
\end{array}
\bigg | \frac{{\beta}^N y^2}{2\epsilon_0} \right),
\label{A:pv}
\end{equation}
with $c = ( 2/\pi \epsilon_0 \alpha^{N})^{1/2}/[K_{p}( \omega)]^N$.

It follows from Eq.~(\ref{NF:18})  that the mean of $f_N(\epsilon_N)$ is obtained by setting $s=2$ in (\ref{NF:21}):
\begin{align}
\langle \epsilon_N \rangle 	&=   \epsilon_0\left[\frac{\alpha K_{p+1}(\omega)}{K_{p}(\omega)}\right]^N,\label{SI18}
\end{align}
which implies 
\begin{align}
\frac{\langle \epsilon_N \rangle }{\langle \epsilon_{N-1} \rangle}	&= \alpha \, \frac{ K_{p+1}(\omega)}{K_{p}(\omega)}.
\end{align}
Now, using $K_\nu(z)\approx \Gamma(\nu) 2^{\nu-1}z^{-\nu}$, for  $z\to 0$, $\nu>0$, it then leads to 
\begin{align}
\frac{\langle \epsilon_N \rangle }{\langle \epsilon_{N-1} \rangle}	&\approx \, \frac{ 2 \alpha \Gamma(p+1)}{\omega\Gamma(p)}=\frac{2\alpha p}{\omega}=1-\alpha^2, \qquad \alpha\to 0.
\label{eq:Amean}
\end{align}
Recursive application of this relation yields
\begin{align}
\langle \epsilon_N \rangle \approx (1-\alpha^2)^N\epsilon_0\approx (1-N\alpha^2)\epsilon_0.
\end{align}

We lastly remark that the novel transcendent $R$-function, which emerges from our $N$-scale intermittency model, seems to have never been previously considered in the literature.

\section{Properties of the $R$-function}

The general usefulness of the $R$-function representation arises from a number of identities that can be derived from extensions of related identities of the Meijer-$G$ function. Therefore, we give below a short list of some general properties of the $R$-function.

	\begin{itemize}
		\item \textit{Mellin transform.}
		\begin{equation}
		\displaystyle \int\limits^{\infty}_{0} dx  \, x^{s-1}  R_{p,q}^{m,n} \left(\begin{matrix} \pmb{a},\pmb{A} \\
		\pmb{b},\pmb{B}
		\end{matrix} \bigg| \alpha\,x \right)   = \alpha^{-s} \widetilde{R}_{p,q}^{m,n} \left(\begin{matrix} \pmb{a},\pmb{A} \\
		\pmb{b},\pmb{B}
		\end{matrix} \bigg| s\right)
		\label{NF:33}
		\end{equation}
		\item \textit{Argument inversion.}
		\begin{equation}
		R_{p,q}^{m,n} \left(\begin{matrix} \pmb{a},\pmb{A} \\
		\pmb{b},\pmb{B}
		\end{matrix} \bigg| x \right) = R_{q,p}^{n,m} \left(\begin{matrix} \pmb{1}-\pmb{b},\pmb{B} \\
		\pmb{1}-\pmb{a},\pmb{A}
		\end{matrix} \bigg| \frac{1}{x} \right)
		\label{NF:34}
		\end{equation}
		\item \textit{Power absorption.}
		\begin{equation}
		x^{\sigma}\,R_{p,q}^{m,n} \left(\begin{matrix} \pmb{a},\pmb{A} \\
		\pmb{b},\pmb{B}
		\end{matrix} \bigg| x \right) = \frac{\displaystyle \prod_{j=1}^{q} B^{\sigma}_j }{\displaystyle \prod_{k=1}^{p} A^{\sigma}_k }  R_{p,q}^{m,n} \left(\begin{matrix} \sigma\pmb{1}+\pmb{a},\pmb{A} \\
		\sigma\pmb{1}+\pmb{b},\pmb{B}
		\end{matrix} \bigg| x \right)
		\label{NF:35}
		\end{equation}	
		\item \textit{Integral involving the product of two R-functions.}
		\begin{equation}
		\displaystyle \int\limits^{\infty}_{0}   R_{n,m}^{m,n} \left(\begin{matrix} \pmb{a},\pmb{A} \\
		\pmb{b},\pmb{B}
		\end{matrix} \bigg| \xi \,x \right)  \, R_{t,r}^{r,t} \left(\begin{matrix} \pmb{c},\pmb{C} \\
		\pmb{d},\pmb{D}
		\end{matrix} \bigg| \eta \,x \right)\, dx  = \frac{1}{\eta}  \frac{\displaystyle \prod_{j=1}^{r} D_j }{\displaystyle \prod_{j=1}^{t} C_j }    {R}_{n+r,m+t}^{m+t,n+r} \left(\begin{matrix}( \pmb{a},-\pmb{d}),(\pmb{A},\pmb{D}) \\
		(\pmb{b},-\pmb{c}),(\pmb{B},\pmb{C})
		\end{matrix} \bigg| \frac{\xi}{\eta}\right)
		\label{NF:36}
		\end{equation}
		
		
	\end{itemize}



\section{Numerical Procedure}

We now provide further details on the numerical procedure to apply our theoretical formalism to the analysis of general (i.e., either numerical or experimental) turbulence data. 

The first step is to determine the optimal window size $M$ to compute the background series of variance estimators  $\left\{\epsilon(k)\right\}$ built from the dataset as described in Section~\ref{sec:data}. 
This is done simultaneously to the fitting of the asymmetry parameter $\mu$.
%

The general idea is to search for the optimal pair $(M,\mu)$ that yields the best agreement between the distribution computed numerically from Eq.~(\ref{eq:P1e}), using the empirical density $f(\epsilon)$, and the empirical distribution of velocity increments. 
In practice, we compute the integral in Eq.~(\ref{eq:P1e}) as a Monte Carlo sum, 
\begin{align}
P(\delta v_r)=\int_0^\infty P(\delta v_r|\epsilon)f(\epsilon) d\epsilon \approx \frac{1}{N_M}\sum_{i=1}^{N_M} \frac{1}{\sqrt{2\pi \epsilon_i}}\, \exp\left\{-{\frac{\left[\delta v_r-\mu\left(\epsilon_i-\langle\epsilon\rangle\right)\right]^2}{2\epsilon_i}}\right\},
%
%
\end{align}
where $\langle\epsilon\rangle=\sum_i \epsilon_i/N_M$ and $N_M=N_v-M$ is the number of windows of size $M$.
If this step is successful  
then one guarantees that a proper modeling of the background density will lead to a good theoretical description of the increments distribution, as described below.  
%
%

We therefore note that the window size $M$ is not a free parameter in the usual sense, but it rather represents an internal length scale that needs to be obtained from the data. Other methods to estimate $M$ for Gaussians with {\it zero} mean have been proposed, e.g., in Refs.~[30-32], but they do {\it not} apply to our case since our conditional Gaussians have nonzero mean, and so it was necessary to find both $M$ and $\mu$ simultaneously.

The next step is to compute the background distribution of the variance series for the optimal value of $M$ and proceed to the fitting of the theoretical prediction, Eq.~(\ref{rn}). Through the Mellin transform formula (\ref{NF:18}) with $s=2$, yielding Eq.~(\ref{SI18}), we can relate the $\varepsilon_0$ parameter to the first statistical moment  of the distribution (\ref{sm:rn}), which is measured from the variance series, and the parameters $\alpha$ and $\beta$. 
This means that $\varepsilon_0$ is not a free parameter, so that the only two free parameters in (\ref{sm:rn}) are $\alpha$ and $\beta$.
These two parameters are then fitted using 
the value of $N$ estimated according to the description in Section~\ref{sec:data}. (For comparison, we also analyze fits for other values of $N$; see main text.)

To perform the fit to Eq.~(\ref{sm:rn}), we must calculate the $R$-function. We note that for~$N$ from 1 up to 6 the multiple integral (\ref{sm:compos}) may be the most efficient way. As mentioned in Section~\ref{sec:theo}, the $N=1$ case is a generalized hyperbolic distribution. Interestingly, the case $N=2$ also allows for an exact integration, and, in fact, for every two new hierarchy levels---and hence two additional integrals in (\ref{sm:compos})---one integral can be executed exactly, reducing at least by  half the number of integrals to be computed numerically.

On the other hand, it is also possible to compute numerically the complex integral~(\ref{NF:30}). 
In this sense, a striking fact is that the aforementioned generalization of the Meijer-$G$ function through the substitution of the gamma functions $\Gamma(\nu)$ by the Bessel functions $K_\nu(x)$ in the Mellin-Barnes integral, Eqs.~(\ref{NF:31}) and~(\ref{NF:32}), greatly simplifies the structure of poles of the integrand. 
Regarding the index $\nu$, the Bessel function for a fixed $x>0$ has a pole only at infinity, and decays to zero for $\nu=c\pm i\infty$. Thus, any vertical contour in the complex plane satisfies the conditions of the Mellin inversion theorem and is suitable for the computation. 
The function grows very rapidly away from $\nu=0$, developing strong oscillations in the real and imaginary parts, which led us to choose a contour that passes through $\nu=0$ in the real line to attain fast numerical convergence. 
For a purely imaginary $\nu$  the function $K_{\nu}(x)$ is real for $x>0$, so that for a single $K$-function the integral in Eq.~(\ref{NF:31}) is real. 
For a product of $K$-functions with different indexes, which happens for any $N>1$, the contour should pass as close as possible to the zeros of these indexes to provide convergence and stability.

Lastly, with all parameters in hand, we 
plot the model prediction for the distribution of velocity increments, which depends on another $R$-function, as given by Eq.~(\ref{pv}), and compare with the one from the original empirical turbulence data.

\bibliographystyle{apsrev4-1}


\end{document}